\documentstyle[epsfig]{mn}

\newcommand{\apj}{ApJ}
\newcommand{\mnras}{MNRAS}
\newcommand{\apjs}{ApJSupp}
\newcommand{\apjl}{ApJ (Letters)}
\newcommand{\aj}{AJ}

\newcommand{\pasp}{PASP}

\title[Star-formation History of HDF Galaxies]{The Star Formation
History of the Hubble Sequence: Spatially Resolved Colour Distributions
of Intermediate Redshift Galaxies in the Hubble Deep Field}

\author[Abraham et al.]{
  R. G. Abraham$^{1,2}$, 
  R. S. Ellis$^2$, 
  A. C. Fabian$^2$, 
  N. R. Tanvir$^2$, and
  K. Glazebrook$^3$\\
  $^1$Royal Greenwich Observatory, Madingley Road, Cambridge, CB3 0EZ\\ 
  $^2$Institute of Astronomy, University of Cambridge, Madingley Road, 
  Cambridge CB3 OHA\\
  $^3$Anglo-Australian Observatory, P. O. Box 296, Epping, NSW 2121, 
  Australia\\
} 

\date{Received:\ \ \ Accepted: }

\pagerange{\pageref{firstpage}--\pageref{lastpage}}%

\begin{document}

\maketitle

\label{firstpage}

\begin{abstract} We analyse the spatially resolved colours of distant galaxies
of known redshift in the Hubble Deep Field, using a new technique based on
matching resolved four-band colour data to the predictions of evolutionary
synthesis models. Given some simplifying assumptions we demonstrate how our
technique is capable of probing the evolutionary history of high redshift
systems noting the specific advantage of observing galaxies at an epoch closer
to the time of their formation.  We quantify the relative age, dispersion in
age, ongoing star-formation rate, and star-formation history of distinct
components. We explicitly test for the presence of dust and quantify its
effect on our conclusions. To demonstrate the potential of the method, we
study the near-complete sample of 32 $I_{814}<$21.9 mag galaxies with
$\overline{z}\sim$0.5 studied by Bouwens et al (1997). The dispersion of the
internal colours of a sample of 0.4$<z<$1 early-type field galaxies in the HDF
indicates that $\sim$ 40\% [4/11] show evidence of star formation which must
have occurred within the past third of their ages at the epoch of observation.
This result contrasts with that derived for HST-selected ellipticals in
distant rich clusters and is largely independent of assumptions with regard to
metallicity.  For a sample of well-defined spirals, we similarly exploit the
dispersion in colour to analyse the relative histories of bulge and disc
stars, in order to resolve the current controversy regarding the ages of
galactic bulges.  Dust and metallicity gradients are ruled out as major
contributors to the colour dispersions we observe in these systems.  The
median ages of bulge stars are found to be signicantly older than those in
galactic discs, and exhibit markedly different star-formation histories.  This
result is inconsistent with a secular growth of bulges from disc
instabilities, but consistent with gradual disc formation by accretion of gas
onto bulges, as predicted by hierarchical theories. We extend our technique in
order to discuss the star formation history of the entire Bouwens et al sample
in the context of earlier studies concerned with global star formation
histories. Finally, we consider how to extend our method using near-infrared
data and to deeper samples.  
\end{abstract}


\section{INTRODUCTION}

The public availability of the Hubble Deep Field (HDF, Williams et al 1996)
has led to a surge of progress in our understanding of the faint galaxy
population. The flatter slope ($d\,log\,N\,/\,dm$) of the optical counts
beyond $B\simeq$25 mag found from earlier deep ground-based studies (Lilly et
al 1991, Metcalfe et al 1995, Smail et al 1995) has been confirmed from the
HDF data. This flat slope, when combined with the relative paucity of
$U$-dropouts (which confirms that most of even the faintest optical sources
have $z<$3 [Guhathakurta et al 1990, Lanzetta et al 1996, Connolly et al
1997]), implies the bulk of the extragalactic background light must have been
produced at a relatively low redshift (Ellis 1997).

The HDF has also contributed much to our understanding of distant galaxy
morphology. Various authors (Abraham et al 1996a; Mobasher et al 1996; Odewahn
et al 1996; Driver et al 1998) have used the HDF to extend the earlier Medium
Deep Survey morphological source counts (Glazebrook et al 1995; Driver et al
1995; Abraham et al 1996b) to fainter limits, finding a surprisingly high
fraction of the $I_{814}<$25 mag sources cannot be categorised within the
Hubble sequence of regular systems. A key question is the physical nature of
these sources and their present-day counterparts.  Another striking result
from the HDF is the small angular sizes of the faintest sources, implying
physical extents of only 2-4 $h^{-1}$kpc.  In a recent analysis, Bouwens et al
(1997) suggest that such sizes cannot be reconciled with the expected redshift
and surface brightness dimming of typical $z<$1 sources and thus claim
substantial physical growth and merging must have occurred for the galaxy
population during the redshift range 1$<z<$3. The deep HDF data complement
recent morphological studies by Brinchmann et al (1998), Lilly et al (1998)
and Schade et al (1998) who have analysed the joint
morphological-spectroscopic data obtained from a HST imaging survey of a
subset of CFRS (Lilly et al 1995) and LDSS (Ellis et al 1996) redshift
surveys. They find strong evolutionary trends in a population of $z<$1
galaxies with irregular morphology but only modest changes in the abundance of
large regular disc and spheroidal galaxies. It is not clear whether these
trends are consistent with semi-analytical models based on hierarchical growth
(Baugh, Cole \& Frenk 1996). For there to be a peak of star formation activity
at $z\simeq$1-2 (Madau 1997), either the evolutionary behaviour of massive
regular galaxies beyond $z\simeq$1 must change dramatically from that observed
for $z<$1, or perhaps the trends delineated from the optical photometric data
are underestimated because of complications such as dust extinction (Meurer et
al 1997).

Another important aspect of the Hubble Deep Field is the ultra-deep
spectroscopic follow-up survey work undertaken with the Keck telescopes (Cowie
1997, Lowenthal et al 1997, Cohen et al 1996). These surveys have confirmed
the conclusions suggested by the photometric analyses above and raised
confidence in the validity of photometric redshifts estimated from multicolour
photometry alone (Ellis 1997, Hogg et al 1998).

In this paper we wish to explore a new approach to the HDF data, somewhat
related to the photometric redshift technique, but which exploits the true
benefit of HST, namely its ability to {\it resolve} distant galaxies.  Rather
than use the multicolour data to predict redshifts for distant sources, we
instead explore the {\it resolved multicolour data for galaxies of known
redshift} using spectral-synthesis models.  For galaxies with
$\overline{z}\simeq$1, the 4-filter HDF colours are dominated by main sequence
stars whose turn-off ages are less than 4-5 Gyr.  As the evolutionary history
of such stars is well-understood, modulo the usual assumptions concerning the
stellar initial mass function and chemical composition, the multicolour HDF
data can be used to derive relative ages for physically distinct components of
distant galaxies, test for the presence of dust and possibly address the
puzzling nature of the irregular sources which appear with such abundance.

This paper is primarily concerned with an initial application of our technique
to a well-defined population of $I_{814}<21.9$ mag galaxies with redshifts
drawn from Bouwens et al (1997). This sample constitutes the faintest
statistically-complete spectroscopic redshift sample currently available in
the HDF. The use of this sample has the benefits that (a) conclusions drawn
from it can be compared to those of independent studies sampling the same
magnitude and redshift range (e.g. Brinchmann et al 1997); and (b) by
restricting our sample to relatively bright limits and moderate redshifts
($\overline{z}$=0.5), we are working with galaxies that are well-resolved and
can be described within the traditional framework of the Hubble System. The
scientific focus of the present paper is directed towards a better
understanding of two contentious yet fundamental issues central to the origin
of the Hubble classification sequence: the relative ages of bulges and discs,
and the formation history of elliptical galaxies.  We will also use our
synthesis results to place some constraints on the previous appearance of our
galaxy sample.

In our own Galaxy the observational evidence for early bulge formation seems
established, mainly since the main tracers of bulge/halo populations (eg.
globular clusters and RR Lyrae stars) are known to be old.  The notion that
bulges form well before the discs in the first stage of galactic collapse is
the essential component in both the original ELS scenario (Eggen, Lynden-Bell,
\& Sandage 1962) and more modern variants (eg. Carney et al. 1990).
Hierarchical galaxy formation scenarios also require old bulges formed by
mergers (eg. Kauffmann, White, \& Guiderdoni 1993; Baugh, Cole \& Frenk 1996),
with visible discs built-up gradually from gas accreted onto these merger
remnants. Recently, however, the issue of the relative age of the bulge and
disc in extragalactic systems has become controversial.  N-body simulations
indicate that bulges form naturally from bar instabilities in discs (Norman,
Sellwood, \& Hassan 1996; Combes et al. 1990), and recent observations now
seem to indicate that bulges display the morphological and dynamical
characteristics of such a formation scenario (eg. triaxiality or ``peanut''
shapes (Kuijken \& Merrifield 1995), cylindrical rotation (Shaw 1993), and
disky kinematics (Kormendy 1992). On the basis of model fits, de Jong (1996)
concludes that age is likely to be a dominant contributor to color gradients
in local spirals. However extrapolation of colour gradients to the inner parts
of the galaxies suggests that at least the inner discs {\em cannot be
distinguished in terms of colour} from bulges (Balcells \& Peletier 1994; de
Jong 1996).  This is perhaps somewhat surprising given the visual impression
from ``true colour'' HST images of distant galaxies, in which bulges seem
generally redder than disks even at quite small galactocentric radii (van den
Bergh et al. 1996). 

In another recent challenge of what had, hitherto, been regarded as a
well-established viewpoint, Kauffmann et al (1996) and Zepf (1997) have
introduced the possibility that elliptical galaxies are continuously formed
from merging systems as expected in hierarchical models for the growth of
structure, rather than formed at high redshift in a single burst of star
formation (c.f. Baade 1957). The observational evidence supporting the
`traditional' view was based primarily on the small scatter in the
colour-magnitude diagram of cluster ellipticals (Sandage \& Visvanathan 1978,
Bower et al 1992). This test has also been extended to higher redshift cluster
samples with morphological classifications from HST (Ellis et al 1997,
Stanford et al 1998). 

Unfortunately, as Kauffmann (1996) points out, cluster samples are ill-suited
to testing the hierarchical picture as they represent accelerated regions so
far as structure growth is concerned. It is quite possible that most of the
rich cluster ellipticals are truly old as their homogeneously distributed
colours imply. The ultimate test of Kauffmann et al's conjecture lies in the
analysis of ellipticals in field samples. The currently-available samples are
too small for robust results (Schade et al 1998) although Zepf (1997) claims
the absence of red objects at very faint limits in the HDF is consistent with
hierarchical predictions. Glazebrook et al (1998) have also claimed evidence
for younger field ellipticals in data from the Medium Deep Survey, on the
basis of $I-K$ colours.

A plan of the paper follows. In $\S$2 we introduce the principles of our
method for analysing spatially-resolved multicolour data, and outline the
basic assumptions and the uncertainties associated with the technique. In
$\S$3 we illustrate how the method works by analysing multicolour data for 3
HDF galaxies of known redshift.  In $\S$4 we introduce the Bouwens et al
sample and discuss its salient properties in the context of earlier
morphological studies.  We partition the sample into sources of ellipticals
and spirals and explore the history of star formation for both types.
Utilising the statistical completeness of the Bouwens et al sample, we use our
derived star formation rates to calculate contributions to the volume averaged
values and predict the recent past appearance of the various classes.  In
$\S$5 we discuss our method more generally and explore briefly the prospects
for validating our results and extending the work to higher redshift systems.
In $\S$6, we summarise our main results. Unless stated otherwise, throughout
this paper our adopted cosmology is one where \hbox{$H_o=70$ km s$^{-1}$
Mpc$^{-1}$}, $\Omega=0.2$, and $\Lambda=0$.

\section{METHODOLOGY}

In general terms any age-dating of galaxies using broad-band photometry is
affected by numerous complications including the age-metallicity degeneracy,
possible effects of internal dust extinction and variations in the stellar
initial mass function (see Renzini 1995 for a review).  However the crucial
advantage of resolved colour analyses is that they make possible studies of
{\em relative} age and star-formation history of the components within
galaxies, which as we will show are far more robust than absolute ages based
upon integrated colours. Furthermore, in the case of HDF galaxies, the
significant look-back time to the distant sources selected for study means
that age uncertainties have far less of an effect than would be the case in
analysing the history of local stellar populations.  {\em Our approach will
thus be to focus more on the relative age distributions of stellar populations
internal to galaxies of a given morphological type, rather than attempting to
determine absolute ages.}  By weighting component age distributions with their
stellar luminosities we hope to address more general questions such as the
recent ($<$5 Gyr) star formation history of disc and spheroidal components.
This is an important goal since, at $z\simeq$0.5, the previous 5 Gyr spans the
important redshift range $z<$2.5.

To understand how 4-filter colour data can be used to place constraints on
past star formation history, consider first a late-type spiral whose star
formation history can be approximated by a constant star-formation rate.  In
reality the constant SFR is effectively a time-average over the appearance and
disappearance of spatially distinct star forming regions, each of which can be
considered as a simple stellar population with a lifetime (before disruption
or gas depletion) which is short compared to the dynamical timescale of the
galaxy. Over time these young populations gradually mix with older generations
of stars and assimilate into older galactic components.  We will translate
this qualitative picture for galaxy evolution into the quantitative language
of spectral synthesis modelling (Charlot \& Bruzual 1991).

The emergent flux from a composite stellar system, $F_\lambda$, at time $t$
can be represented by the convolution of the spectrum of an evolving
instantaneous starburst, $f_\lambda(x)$ with an assumed star-formation rate
(SFR) function $\Psi(t)$:

\begin{equation}
F_\lambda(t) = \int_0^t \Psi(t-x) f_\lambda(x) dx
\end{equation}

$f_\lambda(t)$ term is known given a stellar initial mass function (IMF) and
metallicity from libraries of template stellar spectra and evolutionary
isochrones. For a given star-formation history an age track can be constructed
on a colour-colour diagram using Equation 1. When treating galaxies as point
sources the integrated colours define a single position on a colour-colour
diagram, and given the uncertainties in the assumed values for $\Psi(t)$, the
IMF and metallicity, unique stellar ages cannot be derived.

Since galaxies are not generally homogenous systems with a single age and
star-formation history, {\em when they are spatially resolved we can expect a
dispersion of points on colour-colour diagrams}. The important point we
explore in this paper is whether the nature of this dispersion can be used to
overcome many of the ambiguities inherent in Equation 1.  For example, the
distribution of pixel by pixel colours within a galaxy image is a record of
the manner in which the stellar population has been assembled (eg. via
numerous small bursts, or by a smaller number of larger bursts). We expect
resolved sites of active star formation to approximate instantaneous
starbursts and therefore $\Psi(t)$ to be a $\delta$-function, effectively
breaking the convolution degeneracy in Equation 1. In practice, therefore, the
distribution of pixel colours should directly trace the shape of $f_\lambda$
for a set of ages.

\subsection{Colour-colour Predictions}

Model colour-colour age tracks for the HDF filter set were computed using the
GISSEL96 spectral synthesis code of Charlot \& Bruzual (1991;1996;1998). For
convenience, in our modelling, we assumed a Salpeter initial mass function
(IMF) with low and high mass cut-offs of 0.1 $M_{\odot}$ and 125 $M_{\odot}$
respectively (this latter cut-off is fixed in the standard GISSEL96
distribution).  We also assumed that the colours of individual pixels can be
modelled using exponential star-formation histories with characteristic
timescale $\tau$, ie. $\Psi(t) = \Psi_o e^{(-t/\tau)}$, which is often used to
model the integrated colours of galaxies (Kennicutt 1991). This allows a
fairly generic parameterization of the star formation history, since as
$\tau\to\infty$ this model approximates constant star formation, while
$\tau\to 0$ approximates an instantaneous burst. This exponential form is
clearly an important assumption, but we would like to emphasize two points.
Firstly, if the assumed form is correct then we expect the dispersion in
galaxy colours to scatter along the predictions of exponential $\Psi(t)$
tracks, which is in fact generally seen in our data. The second point is that
exponential star formation histories for individual pixels {\em do not}
necessarily result in a global exponential star-formation history for the
galaxy considered as a whole (ie. when the total light from all pixels is
superposed), unless the galaxy is coeval with a universal star-formation
history applying to all pixels.

Representative colour-colour tracks for stellar populations seen at various
redshifts with different assumed star formation histories and metallicities
are shown in Figure 1. Arrows at the corners of the panels illustrate the
effects of foreground-screen dust extinction in the host galaxy with
rest-frame value $E_{B-V}=0.1$ mag, assuming Galactic, SMC, and local
starburst extinction curves (Pei 1992, Calzetti 1997). Taken together, the
panels in this figure illustrate how the colour-colour tracks change as the
result of varying the star-formation timescale, dust content, and metallicity
of the models.

The colour vs. age tracks on the colour-colour diagrams, $c(t)$, have four
free parameters: age $t$, star formation history e-folding timescale $\tau$,
metallicity $Z$, and dust extinction $E_{B-V}$ (parameterized in terms of
relative optical extinction in the rest frame), i.e. $c(t)$ is shorthand for
$c(t,\tau,Z,E_{B-V})$.  Prior to fitting pixel-by-pixel colours to these
models, all galaxy images were re-registered to sub-pixel accuracy. This was
accomplished by calculating offsets between the different bands (in the
version 2 stacked images provided by STScI) by cross correlating the images,
and then resampling to remove the subpixel positioning errors with spline
interpolation

For each pixel in the galaxy image, we compute the optimum evolutionary track
using a maximum likelihood estimator ${\cal L}$ that is effectively
least-squares:

\begin{equation}
{\cal L}(t,\tau,Z,E_{B-V}) = \prod_{n=1}^{4} {1\over \sqrt{2\pi}\Delta C_n}
 \exp\left( - {(c_n-C_n)^2\over 2 \Delta C_n\,^2}\right)
\end{equation} 

\noindent where the product is over the four colours, and $[C_n, \Delta C_n]$
are the data colours and errors in the pixel (determined using the formulae in
Williams et al. 1996). Appropriate ranges to maximise the likelihood within
were determined from simple astrophysical constraints. In the next section we
describe these astrophysical constraints, and describe in a general way the
effects of varying the model parameters.

\subsection{Qualitative Properties of the Models}

\begin{figure*}
\centering
\epsfig{figure=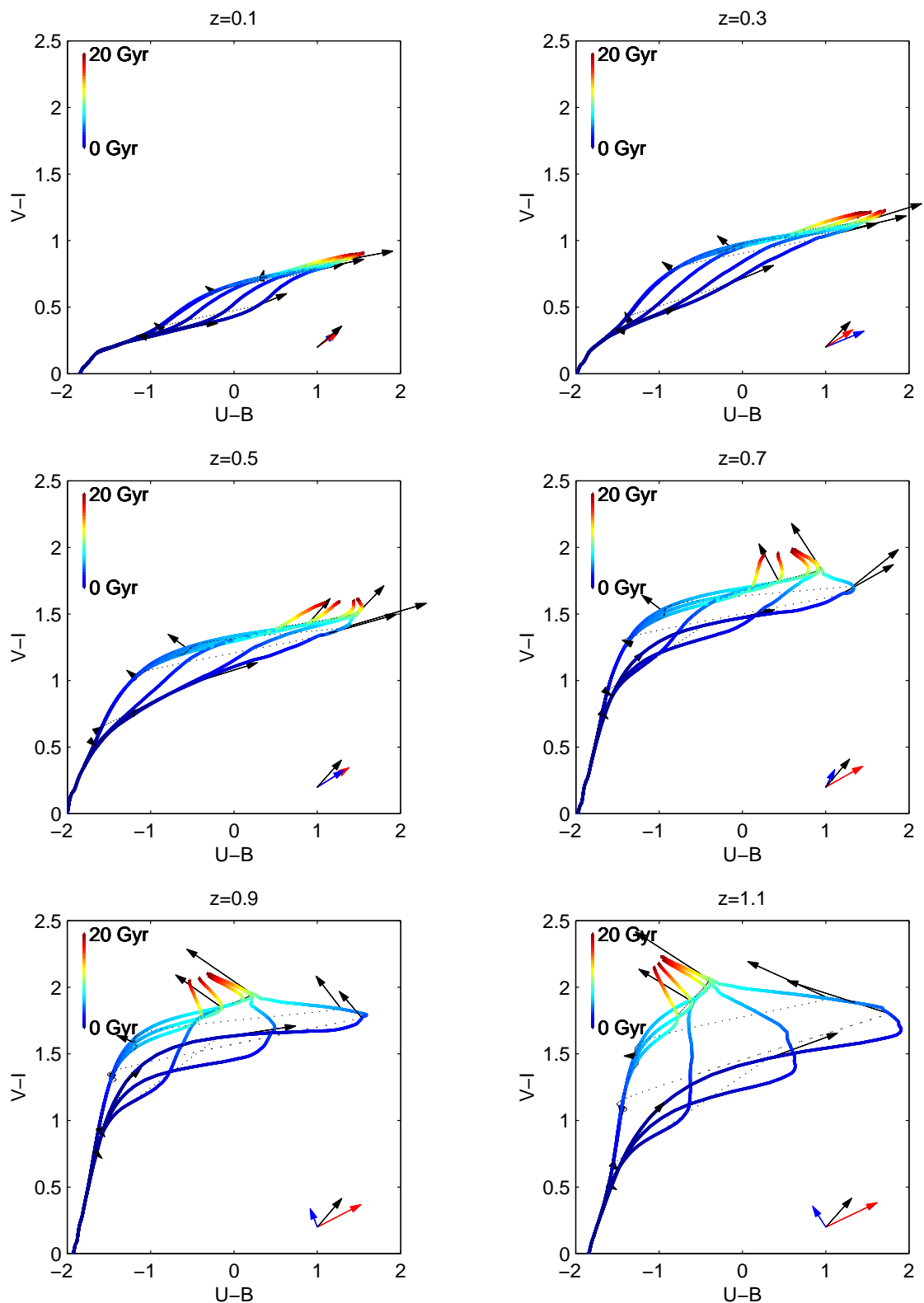,width=15cm}
\caption{\it Age tracks on the HDF $V$-$I$ vs. $U$-$B$ colour colour
plane for model galaxies viewed at various redshifts. The age (time
from the first star formation to the epoch of observation) is indicated
by the coloured bar inthe left corner. The model tracks show
exponentially-decaying star formation histories corresponding to
timescales $\tau$(Gyr)=0.2,0.5,0.8,1.1,1.4,1.7. Dotted lines connect
points at equal ages on the tracks. Arrows on the tracks
indicate the effect of changing metallicities from 30\% solar to
solar.  Arrows with a common origin at the bottom-right of each panel
are vectors indicating the effect of reddening in the host galaxy.
Rest-frame reddening of $E_{B-V}$=0.1 mag for Galactic (blue arrow),
SMC (red arow), and Calzetti (1997) local starburst extinction curves
(black arrow) are shown.  See text for further details. }
\end{figure*}

\subsubsection{Star-formation History}

Unsurprisingly, Figure 1 indicates that the exponential star-formation
timescale $\tau$ (which will be given in units of Gyr throughout this paper)
is the dominant parameter controlling the colour of an evolving stellar
population (Tinsley 1980). The separation between the tracks for the HDF
filter set is most pronounced for redshifts $z>0.3$, so the techniques
described in the present paper are optimized for non-local galaxies when using
the HDF filter set.  As expected, tracks corresponding to large values of
$\tau$ (effectively constant star-formation) predict roughly constant colour
indices at all ages for all redshifts.  Greater trends are seen for evolving
populations with decreasing star-formation timescales.

For models with long and short $\tau$, our sensitivity to the past history
becomes less precise. Clearly the evolution in colour is greatest over
intermediate values of $\tau$, and over ages corresponding to the main
sequence lifetimes sampled in the HDF filter set. In practice, the colours are
not very sensitive to age changes beyond 5 Gyr. For local samples this is a
major deficiency in understanding the history of galaxy formation (Bower et al
1992). However, at a look-back time corresponding to $\overline{z}\simeq$0.5,
the range in stellar ages probed is particularly useful if, as seems likely,
the bulk of the star formation in the Universe occurred at $z\simeq$1.5 (Madau
et al 1997).

At the other extreme, as $\tau$ approaches 100~Myr, the tracks also become
degenerate as $\tau$ becomes comparable with the main-sequence lifetimes of
the massive stars which dominate the rest-frame ultraviolet flux. This is an
important characteristic implying that those populations evolving with
timescales $\tau\sim10^8$ years move along tracks similar in shape to those of
individual star clusters whose formation timescales are $\tau\simeq$10$^{4-5}$
years; in practice the evolution of star clusters can be modeled as
instantaneous starbursts, cf. Meurer et al. 1995. However an effective lower
limit on $\tau$ is imposed by the resolution of HST, since the WFPC-2 pixel
size of $0.1^{\prime\prime}$ corresponds to $\sim1$ kpc at $z\sim1$ and thus
individual star-clusters are not resolved.  The effective limit is given by
the speed at which information can travel across a resolution element. Making
the conservative assumption that information from shocks and winds within a
star-formation complex propagates along the IGM at speeds $< 0.01c$ over
distances $> 1$kpc, in the present paper we adopt a minimum timescale of $\tau
= 5\times10^7$ years. Therefore, for the maximum likelihood analysis in this
paper we explored the range $0.05 < \tau < 3$ Gyr, sampled finely at intervals
$\delta\tau = 0.1$ Gyr.

\subsubsection{Dust}

The three extinction laws adopted in Figure 1 predict markedly different dust
vectors\footnote{ This is partly because local extinctions curves are
differentiated mainly in terms of degree of ultraviolet extinction, and partly
because the 2175\AA~dust feature (prominent in the Galactic extinction curve)
shifts into the bluer HDF filter bandpasses at moderate redshifts.}. Because
of this, a strong variation in galaxian internal colour originating mostly
from dust extinction, as opposed to age, is difficult to rule out in any
individual object. Indeed, it is usually the case that for any individual
galaxy an appropriate combination of attenuation strength and extinction law
can be chosen in order for dust obscuration to mimic the colour variations
predicted by a range in ages.  Fortunately, this is not true for our galaxy
samples considered as a whole because a general ``dust-age degeneracy'' would
require fine-tuning in the choice of extinction law and attenuation strength
as a function of redshift.  We might also assume that any dust present in
distant galaxies would be distributed similarly to that in local systems, at
least for systems with regular appearance. If, as is the case locally, dust is
not distributed uniformly on scales larger than a few kiloparsecs, then for
pixels corresponding to physically large features, scatter in the colours
along the direction of predicted dust vectors might be taken as a sign of
extinction, while uniform colours can be taken as a sign of generally low dust
content.  Since (as will be shown) the scatter in the internal colours of the
galaxies studied in this paper is mostly along the predicted age tracks, dust
is likely to be playing a subsidiary role in defining the internal galaxian
colours.  (The availability of both $U_{300}$ and $B_{450}$ provides an
important constraint given the significant redshifts for most of our galaxies,
and the sharply rising dust extinction in the UV.)

In addition to these general considerations which apply to the sample as a
whole,  we have also chosen to incorporate dust explicitly as a free parameter
in our maximum-likelihood analysis, using the extinction law of Calzetti
(1997).  We chose to consider only the single Calzetti law in our
maximum-likelihood analysis as a compromise between choosing to neglect dust
completely, which we felt to be rather unrealistic in spite of the general
considerations described earlier, and choosing to allow complete freedom in
the choice of extinction law, which would enormously complicate the analysis.
In the future more solid constraints on the dust content (and appropriate
extinction law) may be obtained by augmenting the star-formation histories
inferred from the present analysis with integrated $JHK$ and $H+K-$band fluxes
(Cowie 1997, Dickinson et al 1998) or their resolved equivalents using NICMOS.
Future data may also help to constrain the geometry of the dust extinction.
The present paper assumes extinction can be modelled using a patchy foreground
screen, which we adopted purely for the sake of simplicity, although it should
also be noted that Calzetti (1997) argues that a foreground screen model gives
a somewhat better representation of the extinction in starburst galaxies than
does a mixed dust model.  

\subsubsection{Chemical Enrichment}

The age-metallicity degeneracy that is a common characteristic of
colour-colour diagrams for early-type galaxies and globular clusters is
evident in the tracks shown in Figure 1. In the context of the present paper,
where many of the stellar populations being studied are rather young, the
degeneracy is less troublesome. For young populations studied in the
rest-frame UV, the observable light is dominated by luminous stars at the high
mass end of the IMF. Such stars are less sensitive to metallicity variations
(as seen in Figure 1) and thus, for the first few Gyr, modest metallicity
variations cannot mask a large range in ages.

For ages $t > 3$ Gyr metallicity gradients {\em can} in principle play an
important role in defining the internal colours of galaxies. In addition,
although our synthesis models do incorporate post-MS contributions, it is
possible that a UV excess attributed by us to younger stellar populations
could represent some deficiency in the treatment of post-MS phases as has been
the case for nearby ellipticals (Charlot, Worthey, \& Bressan 1996).  As was
the case with local data, however, the spatial resolution of the UV light is a
key discriminant between young and old stellar populations.

Throughout the remainder of this paper we will explore metallicity variations
by restricting consideration to a fixed set of metallicities appropriate for
local Hubble sequence galaxies: 30\% Solar, Solar, and 150\% Solar.

\subsection{Key Assumptions}

In summary, our methodology is based on three key assumptions: (1) GISSEL96
models with exponential star formation histories accurately model the colours
of individual pixels on galaxy images.  (2) Dust extinction can be
parameterized by patchy foreground screen extinction using the Calzetti (1997)
dust law.  (3) Metallicity lies in the range 30\% Solar - 150\% Solar.

The first assumption can be justified internally from the data given the
qualitatively good agreement between the scatter in pixel-by-pixel colours and
the shapes of exponential star formation history tracks. The agreement between
this dispersion and the models also suggest that dust plays a subsidiary role
in the distribution of galaxian internal colours (at least at the moderate
redshifts explored in this paper), although no attempt will be made to
constrain the detailed form of the dust law from our data. The final
assumption simply corresponds to metallicity lying within a reasonable large
range of values representative of those seen in local Hubble sequence
galaxies.  

\section{Illustrating the Method}

Figures 2--4 show results, based upon the colour-colour tracks introduced in
$\S$2, for three HDF galaxies subjected to our internal colour-colour
analysis.  These examples have been taken from the statistically-complete
sample given in Table 1 of Bouwens et al (1997), and illustrate the types of
conclusions (and their associated uncertainties) we can derive concerning the
past star formation history of resolved high-redshift galaxies. We discuss
these examples individually here, and defer discussion of the entire sample
until $\S$4. Throughout the remainder of this paper objects will be referred
to using the identifications given in the Williams et al. (1996) HDF
catalogue.

\begin{figure*}
\centering
\epsfig{figure=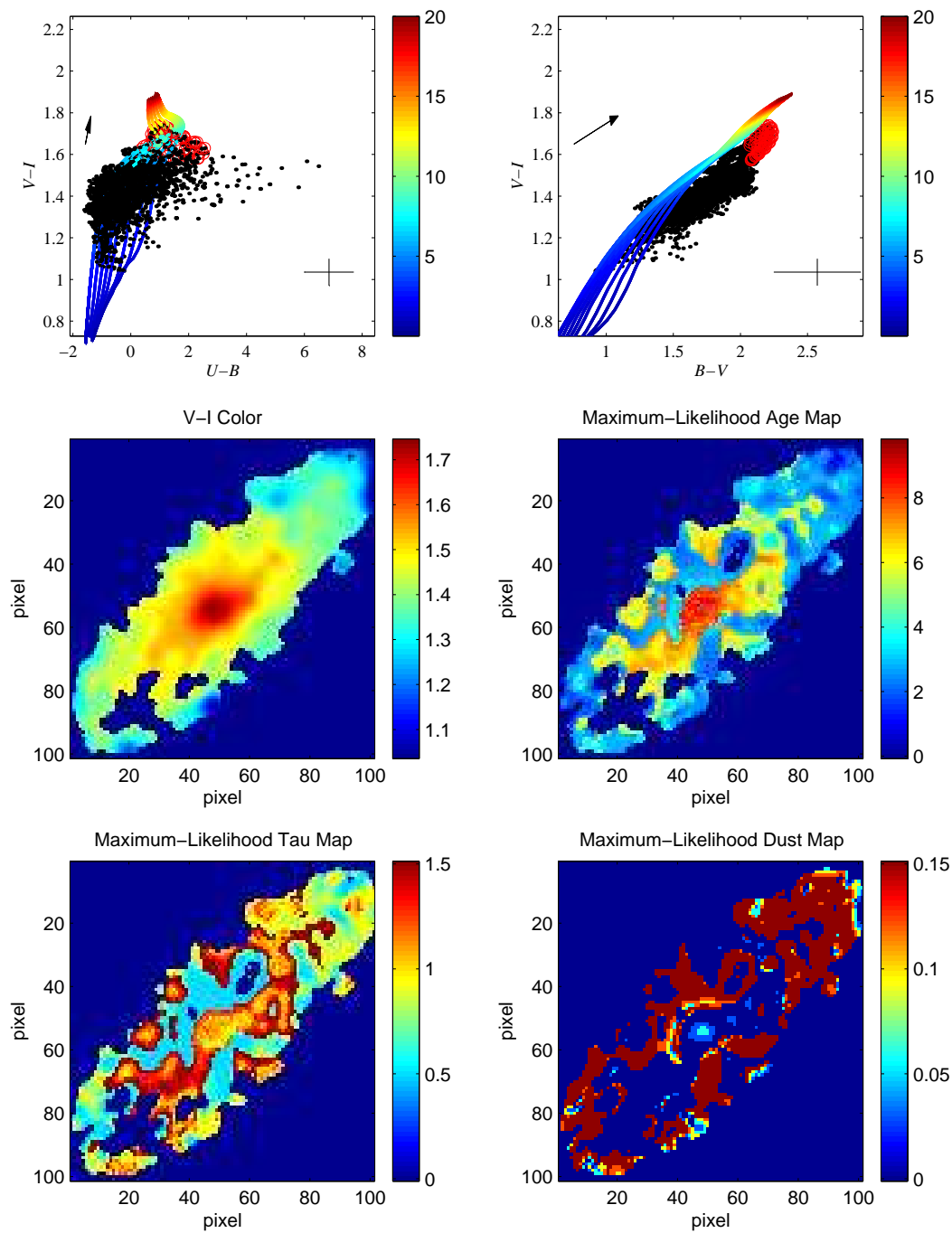,width=15cm}
\caption{\it Illustration of our colour-colour analysis as applied to
a $z$=0.517 spiral (HDF-3.610) taken from the catalogue of Bouwens et al
(1997). Representative {\em dust-free} 
model tracks for solar metallicity are
shown in the top two panels, with points corresponding to
the colours of individual pixels superposed. Red points refer to 
pixels within a 5
pixel radius of the center of the image. The pixel-by-pixels colours in
the 4 bands are compared to the predictions of the spectral synthesis models
for the redshift in question, using a finer model grid than shown
in the top panels, and including the effects of variable dust extinction.
The spatial colour $V-I$ map is shown at middle left, and refers
to an area 4 $\times$ 4 arcsec sampled at the drizzled pixel scale of
0.04 arcsec and limited to contiguous pixels contained within the
$\mu_B$=26.0 mag arcsec$^{-2}$ isophote.
The corresponding age, star
formation timescale, and dust maps are determined using the
maximum-likelihood formulation described in the text. Corresponding
best-fit maps for age, $\tau$, and $E_{B-V}$ are shown in
other panels.}

\end{figure*}

\begin{figure*}
\centering
\epsfig{figure=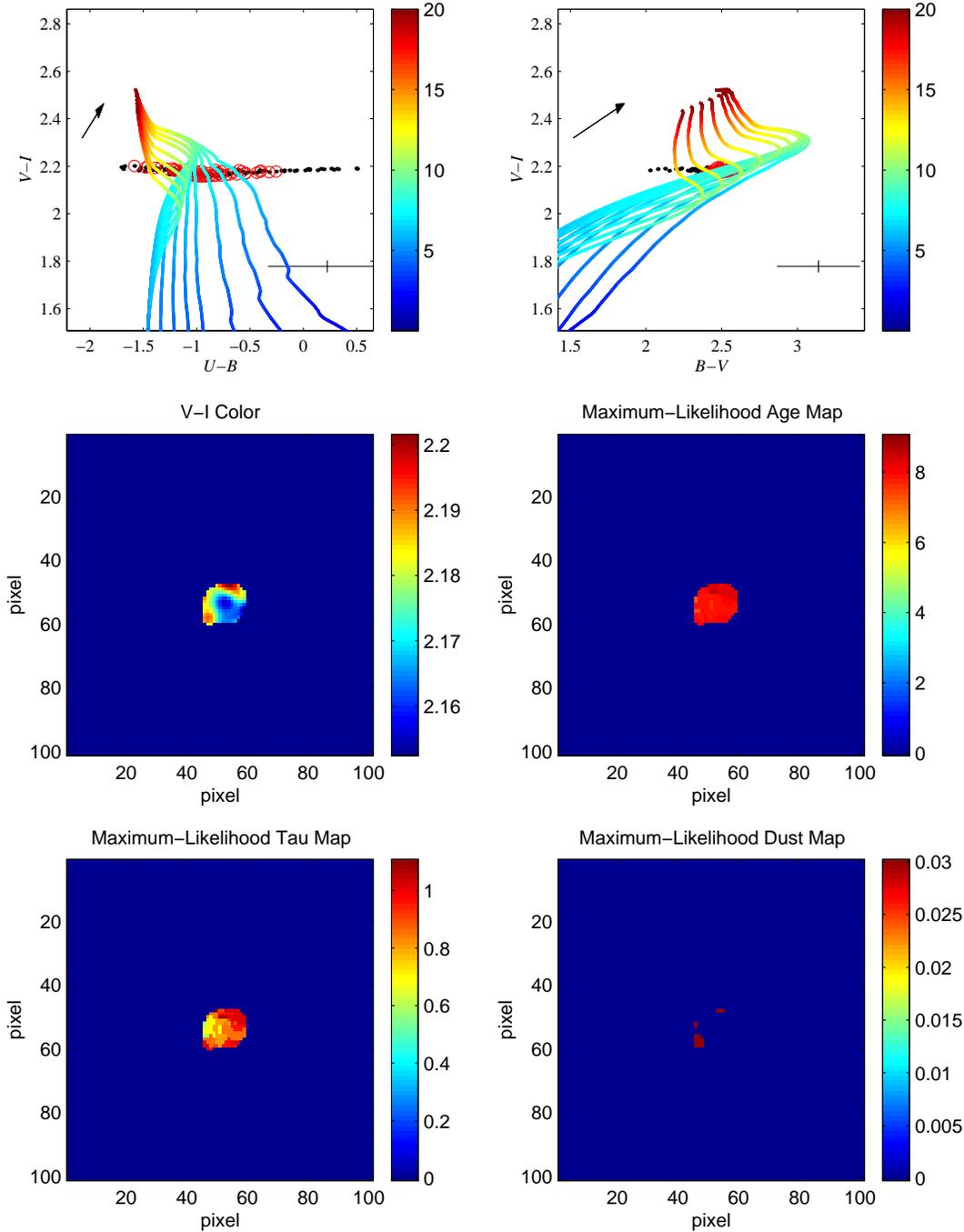,width=15cm}
\caption{\it As for previous figure, except showing a coeval, old 
elliptical
system (4.752) at $z$=1.013.}
\end{figure*}

\begin{figure*}
\centering
\epsfig{figure=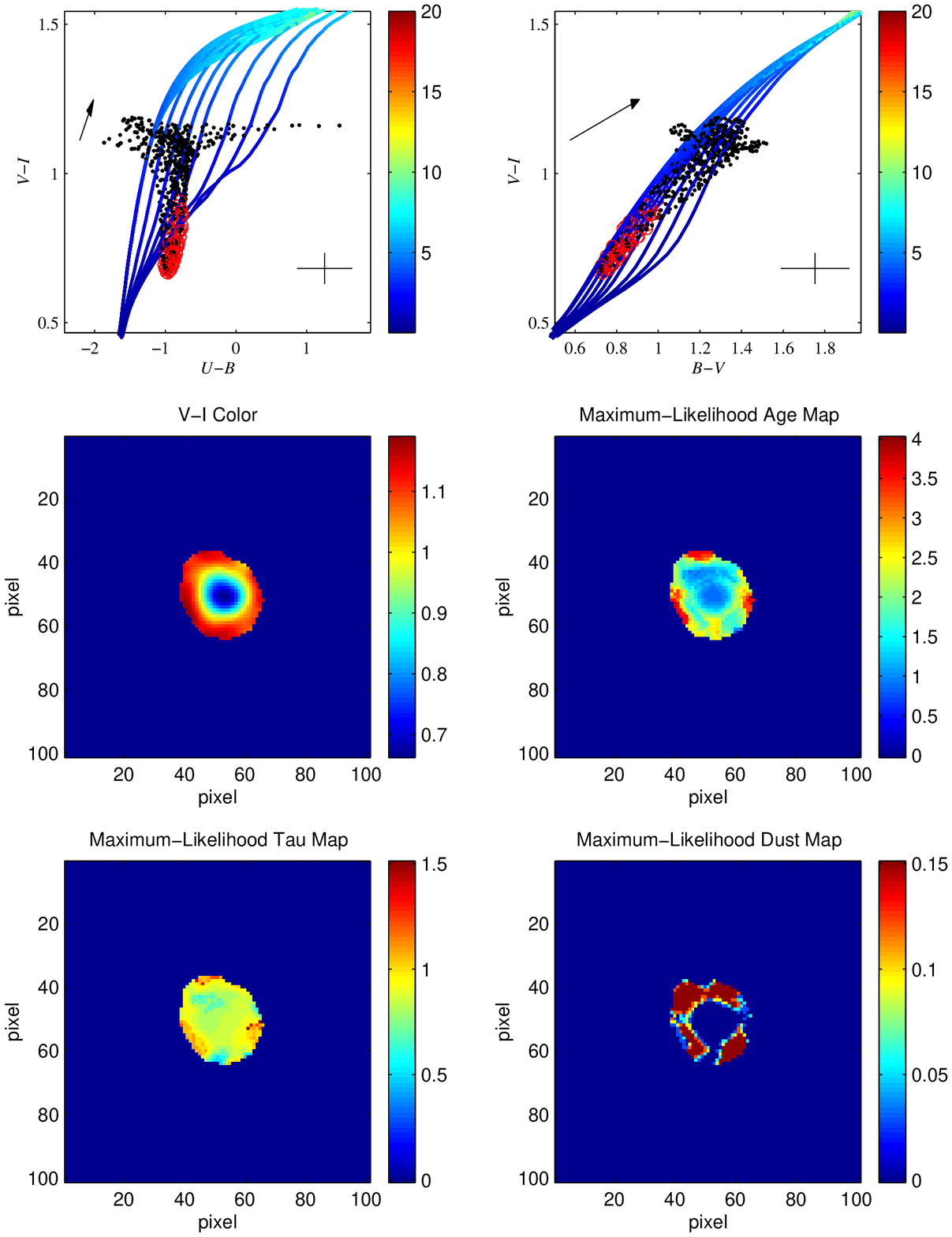,width=15cm}
\caption{\it As for previous figure, except for an early-type system
(HDF-2.264) at $z=0.478$ showing evidence of recent star formation.}
\end{figure*}

\subsection{HDF-3.610 -- a $z$=0.517 spiral}

Figure 2 [middle left] shows the $V-I$ colour image of a bright
($I_{814}$=19.7 mag) spiral at $z$=0.517 in the HDF (HDF-3.610, the
third galaxy in Bouwens et al's Table 1). The galaxy has been isolated
from the background sky by selecting all contiguous pixels within a
surface brightness threshold of $B_{450}$=26.0 mag arcsec$^{-2}$.  An
important consideration in selecting the isophote and the optimum
filter is to ensure a minimum signal/noise per pixel over all
wavebands.  On the one hand, as our eventual sample is $I$-selected, it
might seem appropriate to select the isophote at a longer wavelength.
However, this would bias the pixels selected to those less sensitive to
recent star formation and in many cases there would be inadequate
signal in the shorter wavelength filters. Were the pixels to be
selected according to their $U$-band signal/noise, undue preference
might be given to strong SF regions particularly for the higher
redshift galaxies. We experimented with selection in both the $B$ and
$V$ bands and found no great difference between the results. In most
cases the colour error for each pixel selected according to these
criteria is better than 0.2-0.3 mag.

In the top two panels of Figure 2, colour-colour distribution of pixels
in the two possible planes ($V-I$ vs. $U-B$ and $V-I$ vs. $B-V$) are
compared with the predictions of solar-metallicity, {\em dust-free}
($E_{B-V}=0$) evolutionary tracks. The tracks shown are purely for
illustration -- in the full maximum-likelihood analysis the tracks are
more finely sampled and dust content is a free variable.  Also for
illustrative purposes, the colours of pixels in the central region of
the galaxy have been coded with red points in the colour-colour plots.
For this galaxy the majority of the points lie well within the locus of
the $V-I$ vs $U-B$ colour space mapped by the tracks, although some red
outliers (corresponding to low signal-to-noise points) lie redward of
the tracks. On the other hand the majority of points lie slightly
systematically redward of the predicted $V-I$ vs $B-V$ tracks,
suggestive of mild dust extinction.  A crucial point is that the {\em
dispersion} in colour-colour space is in excellent agreement with the
direction predicted by an age gradient along the colour-colour tracks.
This is a generic feature of the galaxies in the present sample, even
amongst those objects that show small systematic offsets from the model
tracks (most -- but not all -- of which can be accounted for by pockets
of Calzetti extinction law dust obscuration).  \footnote{Our measure of
absolute goodness of fit to a set of model tracks (the mean likelihood
over all pixels) will be described in sections 4.1 and 4.2. The example
spiral described here has a goodness-of-fit that is actually slightly
worse than the mean for the spiral sample}.

Best fitting age, $E_{B-V}$ and $\tau$ maps from the full
maximum-likelihood analysis are shown in Figure 2 as separate 
panels. Parameter ranges are keyed to the vertical colour scale.  From
an examination of Figure 2 it is clear that this $z$=0.517 spiral
has a bulge which is distinctly redder in all colours and considerably
older than that of the disc. More accurately, the optical colours of
the disc reflect ongoing star formation in the past 5 Gyr whereas
there is no evidence that this has been the case for the bulge. It is,
of course, not possible to eliminate the possibility that the disc
also contains an underlying population of red stars coeval with those
of the bulge; its visibility would be difficult to detect given the
abundant population of younger stars. Two points emerge from this
restriction. Foremost, the relative visibility of the disc and bulge
components in the era 1$<z<$2 just prior to the epoch of observations,
will not be affected. Secondly, the amount of mass involved in an
early population could be constrained from near-infrared
luminosities.  

As alluded to earlier, significant ($E_{B-V}\sim 0.1-0.15$) pockets 
of patchy dust do improve the fit
to the disc colours, but dust is absent from the central regions in
this galaxy. In fact the dust seems to be distributed along the inner
parts of weak spiral arms, as is often seen in local galaxies.
For the comparison shown in Figures 2--4 we have assumed
solar metallicity, but as shown in Figure 5 varying metallicity cannot
result in synchronization of the ages in the inner and outer parts of
this system.  The distribution of star formation timescales ($\tau$)
shown in Figures 2--4 requires careful explanation. As the tracks in
the colour-colour plane show, we have greatest sensitivity to $\tau$
for populations of intermediate age. For the bulge light there is
almost complete degeneracy in solving for $\tau$.  Accordingly the
slight mismatch between the model prediction for $B-V$ and the
observations can swing the fit to a high $\tau$ whereas the
convergence of the tracks at these ages would be compatible with any
value.  However even for the shortest star-formation timescales the
age of the bulge is considerably older than that of the disc, ruling
out the possibility that star-formation timescales, rather than age,
is responsible for the redder bulge light. In fact no variation in
star-formation history, metallicity, or dust alters the basic
conclusion that the redder bulge colours are simply due to these
pixels originating from the oldest starlight in the galaxy. However
Figure 5 does show that varying metallicity within reasonable limits
(the excellent agreement between the observed dispersion of colours
along the solar metallicity tracks effectively restricts the
metallicity of the disc light in this case to 75\% $<$ $Z/Z_\odot$
$<$120\% Solar) shifts the absolute ages of the various components
in this system -- confirmation that our technique is best used to
study only relative differences in age. In \S4 we define a simple
statistic which quantifies the relative ages of the inner and outer
portions of all spirals in our sample.

\subsection{HDF-4.752 -- a $z$=1.013 elliptical}

Our second example (Figure 3) is a high redshift elliptical with
$I_{814}$=20.9 mag. The object is remarkably homogeneous in its $B-V$
and $V-I$ colours and consistent with a coeval formation of all
visible starlight. The red spectral energy distribution means that the
$U$ image is extremely faint which leads to a large random error in
the $U$--$B$ colours (except for the brighter central regions coloured 
red in the colour-colour diagrams shown in Fig. 3).  Because the colours of
this galaxy correspond (in effect) to a single point on the
colour-colour diagram, no particular age-track is mapped out, and thus
the system is particularly subject to the well-known age-metallicity
degeneracy that plagues analysis of integrated colours of early type
systems. This point is shown clearly in Figure 5 (b) which shows the
$V-I$ vs. $U-B$ colour plane for models with 30\% solar, solar and
150\% solar metallicities. The key point here is that, although the
absolute age inferred is indeed critically dependent on the adopted
metallicity, two important conclusions remain unchanged: (i) the light
of the galaxy is consistent with a {\it single} stellar age, (ii) the
age is cosmologically significant given the large look-back to this
source.

\subsection{HDF-2.264 -- a $z$=0.478 elliptical}

The final example (Figure 4) consists of an elliptical with
$I_{814}$=21.8 mag at $z$=0.478. In this case there is an obvious
spread in the internal colours in this system. We see the same spread
in $U$-$B$ colours as in Figure 3 corresponding to the lower signal to
noise in the $U$ band for faint sources. However, the nuclear region
of the galaxy is significantly bluer than the outer galaxian light,
consistent with a younger stellar age (this object might be considered
an example of a ``blue nucleated galaxy'' (Schade et al. 1995), except
that
it is not morphologically peculiar). This remarkable conclusion
holds even when the maximum likelihood analysis attempts to minimize
the age spread in this system by invoking a conveniently arranged
annulus of dust.  The most straightforward interpretation of the
colour data is that this E/S0 is {\it not} a simple stellar system and
that some fraction of the starlight was produced within 2 Gyr of the
epoch of observation.  In the next section we will quantify the
frequency of occurrence of cases with such star-formation histories.

\begin{figure*}
\centering
\epsfig{figure=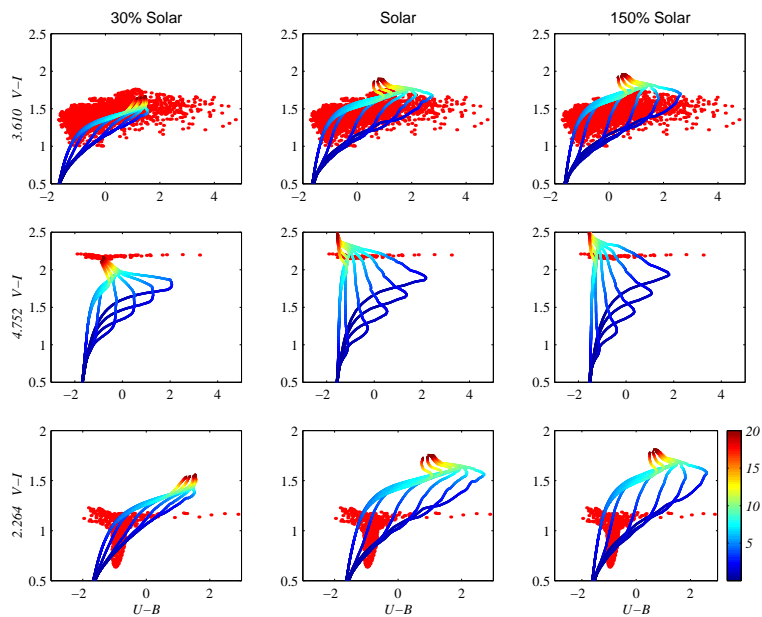,width=110mm}
\caption{\it The sensitivity of the evolutionary tracks to the 
adopted metallicity in terms of the $V$-$I$ versus $B$-$V$ colour plane for 
the 3 galaxies presented in Figures 2--4. A range in $\tau$ from 0.1 Gyr
to 1.6 Gyr, in steps of 0.2 Gyr, is shown. Columns from left to right
correspond to tracks with metallicities of 30\% solar, solar, and 150\% solar.}
\end{figure*}

\section{Analysis of the Bouwens et al HDF Sample}

In order to apply the methodology illustrated above to a representative
sample, we now consider the $I_{814}<$21.9 mag galaxies in the Bouwens et al
(1997) sample, as this constitutes the faintest statistically-complete sample
in the HDF with redshift information.  Because the model tracks are
considerably more degenerate at low redshifts when using the  HDF filter set,
we will not consider the three galaxies with $z<0.3$ in the Bouwens et al
list. Thus our sample consists of 28 galaxies with $0.318<z<1.015$.

Firstly, we consider the distribution of galaxy properties within the broad
morphological categories (E/S0: Sabc: Irr/pec/merger) adopted initially by the
Medium Deep Survey team (Glazebrook et al 1995).  A puzzling feature of the
Bouwens et al catalogue is that only a small number (3/31) are apparently
`Irr/pec/merger' whereas, within the entire MDS survey to the same magnitude
limit, this fraction is close to 30\%. In order to check this result and to
facilitate comparison with the survey of Brinchmann et al (1997), the Bouwen
et al HST images have been reclassified by one of us (RSE) and absolute
magnitudes in $M_{B(AB)}$, recomputed using K-corrections corresponding to the
mean luminosity-weighted age and star-formation history from our model fits.
The revised properties of the Bouwens et al sample are listed in Table 1.

We find good agreement between our absolute magnitudes and those of Bouwens et
al. However the morphological classifications of a number of systems are
rather different.  Three objects categorised as spirals by Bouwens et al
(HDF-3.400,HDF-3.143, and HDF-4.950, corresponding to \#'s 16, 27, 30 in
Bouwens' table 1) have been classed by us as `Irr/pec/merger' and a fourth
(HDF-4.402, \#22 in Bouwens et al's table 1) is a marginal case. Although a
somewhat subjective correction, the visual classification doubles the number
in this category reducing but not eliminating the discrepancy with the MDS
counts. However, examination of the Irr number counts within the HDF itself to
$I\simeq$22 (Abraham et al 1996) does indicate a marginal deficiency with
respect to the more extensive MDS counts which we presume arises simply from
small number statistics.

\begin{table*}
\centering
\begin{minipage}{140mm}
\caption{The $z>0.3$ Sample from Bouwens et al. (1998). Primary object identifications
are correspond to entries in the Williams et al. (1996) catalogue, indexed against entries in Bouwens
et al's table~1. The three galaxies with $z<0.3$ are not
tabulated.}
\begin{tabular}{@{}cccccc}
\hline
 Williams ID  &  Bouwens ID  &  $z$  &  $m_{I814}$ & $M_{B450}$ & RSE Class \\
\hline\hline
 3.610  &   3  & 0.517 &  19.85 & -21.59 &   S \\
 2.121  &   4  & 0.475 &  20.03 & -21.19 &   Epec \\
 3.386  &   5  & 0.474 &  20.27 & -21.25 &   Spec/Pec \\
 4.656  &   6  & 0.454 &  20.29 & -21.10 &   S \\
 4.744  &   8  & 0.764 &  20.41 & -22.39 &   E \\
 3.350  &   9  & 0.642 &  20.48 & -21.81 &   S \\
 4.795  &  10  & 0.432 &  20.54 & -20.65 &   S \\
 4.241  &  11  & 0.318 &  20.33 & -20.21 &   S0 \\
 2.652  &  12  & 0.557 &  20.66 & -21.20 &   Pec \\
 4.550  &  13  & 1.012 &  20.91 & -23.03 &   S \\
 3.534  &  14  & 0.319 &  20.94 & -19.58 &   S \\
 2.251  &  15  & 0.960 &  20.86 & -22.92 &   E \\
 3.400  &  16  & 0.473 &  20.94 & -20.64 &   Pec \\
 3.790  &  17  & 0.562 &  20.93 & -20.82 &   E \\
 3.321  &  18  & 0.677 &  20.92 & -21.44 &   S0 \\
 3.264  &  19  & 0.475 &  21.03 & -20.37 &   S \\
 4.752  &  20  & 1.013 &  20.88 & -23.37 &   E \\
 3.659  &  21  & 0.299 &  21.07 & -18.99 &   S \\
 4.402  &  22  & 0.555 &  20.89 & -21.06 &   Spec \\
 4.493  &  23  & 0.847 &  21.17 & -22.16 &   S0 \\
 2.514  &  24  & 0.752 &  21.34 & -21.44 &   Sa \\
 3.486  &  25  & 0.790 &  21.46 & -21.50 &   Sp \\
 4.471  &  26  & 0.504 &  21.51 & -19.89 &   E \\
 3.143  &  27  & 0.475 &  21.50 & -19.93 &   Pec \\
 2.264  &  28  & 0.478 &  21.42 & -20.15 &   E/S0 \\
 4.402  &  29  & 0.556 &  20.29 & -21.60 &   Pec/Mrg \\
 4.950  &  30  & 0.608 &  21.72 & -20.46 &   Irr \\
 4.928  &  31  & 1.015 &  21.75 & -22.30 &   E \\
\hline
\end{tabular}
\end{minipage}
\end{table*}

\subsection{Spheroidal Galaxies}

We begin by addressing the uniformity (or otherwise) and the star formation
history of the 12 E/S0 galaxies in the Bouwens et al sample.  We note that 2
of the ellipticals (HDF-2.251 and HDF-4.752, \#'s 15, 20 in Bouwens et al's
table 1) are strong radio sources (Fomalont et al 1997). We discount the first
object in Bouwen's table because of its low redshift ($z$=0.089), and also
HDF-2.514 which is classed as an elliptical by Bouwens et al. but as an Sa by
us.

The key question here, much debated in the recent literature, is whether field
ellipticals show evidence of recent star formation, (as expected if they are
continuously formed from merging systems (Kauffmann et al 1996) or are coeval
in the sense of having formed most of their stars via a single burst at high
redshift. The question can be addressed observationally in a number of ways.
The only morphological data from HST that has been discussed in this context
is that of Schade et al (1998) who illustrate the dangers of relying purely on
optical colours to select passively-evolving systems at various redshifts.
Their sample of E/S0s from the CFRS/LDSS survey is too small to make strong
arguments concerning absolute number densities at various redshifts. However,
in terms of surface photometry and colour evolution, their analysis showed
that most (morphologically-selected) spheroidals are following the passive
evolutionary track albeit with more scatter from galaxy-galaxy than was found
for cluster ellipticals (Ellis et al 1997).

Our technique offers a fresh look at this problem since we are able to
consider the {\it internal} homogeneity in the multicolour photometry.  Figure
6 shows three representative moderate $z$ E/S0 galaxies drawn from this
sample. It appears that HDF-3.790 and HDF-4.471 are consistent with an old
stellar population; our analysis indicates no ongoing SF activity in the 2 Gyr
prior to $z\simeq$0.5. The bulk of the SF in such galaxies must have occurred
at much earlier times. However, HDF-2.264 shows a much more recent history of
SF activity with much of the activity in the previous 0-2 Gyr.

%%% New bit added by RSE

We now wish to examine all of the E/S0 galaxies in this way. Recognising the
Bouwens et al sample is obviously small and covers a range in redshift, we
seek a simple characterisation of the difference between the two histories
discussed above. Foremost it would be helpful to determine whether there is a
greater dispersion in the star formation rate of {\it field} ellipticals
compared to their {\it cluster} counterparts. Rather than interpret the data
in the context of hierarchical models (Baugh et al 1996, Kauffmann 1996) we
first seek an empirical comparison between two such samples.

For each spheroidal galaxy in the Bouwens sample, our methodology enables us
to measure the fraction of the stars formed in the last third of the galaxies'
lifetime, $\Psi_{1/3}$, where we define the start of the galaxy's lifetime as
the age of the oldest pixel in the image. This is given for various
metallicities and dust assumptions in Table 2.

\begin{table*}
\centering
\begin{minipage}{150mm}
\caption{Early Type Galaxies: Fractional Star-formation in Last 1/3 of Lifetime}
\begin{tabular}{@{}cccccccccccc}
\hline
      &     &   \multicolumn{3}{c}{solar + dust} &   \multicolumn{3}{c}{30\% solar + dust} & \multicolumn{3}{c}{solar + no dust} & \\
 ID   &  $z$  &  $\Psi_{1/3}$ &  $\langle {\cal L} \rangle$\footnote{Mean likelihood for all pixels.}  & $\langle E_{B-V}\rangle$\footnote{Mean rest-frame relative $B - V$ dust reddening for all pixels.} &  $\Psi_{1/3}$ &  $\langle {\cal L} \rangle$  & $\langle E_{B-V}\rangle$ &  $\Psi_{1/3}$ &  $\langle {\cal L} \rangle$  & $\langle E_{B-V}\rangle$ &  comment \\
\hline\hline
2.121  &     0.475 &     0.012 &     0.601  &     0.020 &     0.000  &     0.678 &     0.027  &     0.006  &       0.596    &         0.000  &  E  (coeval) \\
4.744  &     0.764 &     0.002 &     0.584  &     0.001 &     0.000  &     0.822 &     0.009  &     0.001  &       0.583    &         0.000  &  E  (coeval) \\
4.241  &     0.318 &     0.230 &     0.743  &     0.017 &     0.212  &     0.650 &     0.021  &     0.208  &       0.739    &         0.000  &  S0         \\
2.251  &     0.960 &     0.092 &     0.791  &     0.098 &     0.053  &     0.777 &     0.061  &     0.059  &       0.709    &         0.000  &  E         \\
3.790  &     0.562 &     0.019 &     0.312  &     0.003 &     0.002  &     0.611 &     0.004  &     0.016  &       0.312    &         0.000  &  E  (coeval) \\
3.321  &     0.677 &     0.003 &     0.597  &     0.006 &     0.000  &     0.842 &     0.029  &     0.002  &       0.592    &         0.000  &  S0  (coeval) \\
4.752  &     1.013 &     0.004 &     0.920  &     0.003 &     0.000  &     0.748 &     0.007  &     0.006  &       0.918    &         0.000  &  E  (coeval) \\
4.493  &     0.847 &     0.001 &     0.885  &     0.022 &     0.000  &     0.827 &     0.039  &     0.001  &       0.858    &         0.000  &  S0  (coeval) \\
4.471  &     0.504 &     0.005 &     0.400  &     0.000 &     0.001  &     0.583 &     0.013  &     0.005  &       0.400    &         0.000  &  E  (coeval) \\
2.264  &     0.478 &     0.365 &     0.807  &     0.070 &     0.278  &     0.733 &     0.047  &     0.318  &       0.757    &         0.000  &  E/S0         \\
4.928  &     1.015 &     0.058 &     0.350  &     0.144 &     0.036  &     0.709 &     0.121  &     0.040  &       0.192    &         0.000  &  E         \\
\hline
\end{tabular}
\end{minipage}
\end{table*}

$\Psi_{1/3}$ is a useful physical measure for several reasons. Most
importantly, as Table 2 shows, it is fairly robust to changes in the assumed
metallicity. Whereas absolute ages depend crucially on metallicity, the {\it
relative} ages of young and old stellar populations are largely unaffected. It
is also a quantity that can readily be predicted in those hierarchical models
which assume E/S0 galaxies are continuously produced via merging of gaseous
dark matter halos (Kauffmann et al 1996). In the case where ellipticals form
by monolithic collapse at high redshift ($z_F>$3) one expects
$\Psi_{1/3}\simeq$0 for all objects.  The critical question then is the extent
to which non-zero values in Table 2 are informative and, in particular, differ
from what is known about cluster ellipticals at high redshift. 

Resolved multi-colour data of comparable quality does not exist for any sample
of distant cluster ellipticals. Given some assumptions, the tight photometric
scatter in the rest-frame ultraviolet colours of cluster ellipticals  (Ellis
et al 1997, Stanford et al 1998) can be used to derive upper limits to recent
star formation in those galaxies.  In what follows, we examine a preliminary
comparison of cluster and field star formation histories.

Following Bower et al (1992), we assume that the scatter in rest-frame $U-V$
colour originates in secondary bursts of star formation in the last few Gyr
prior to the epoch of observation. For the cluster ellipticals analysed by
Ellis et al, whose mean redshift is $<z>$=0.55, the observed rms scatter for
the E/S0 population is $\delta(U-V)$=0.06 mag $\pm$0.01. If this scatter is
produced by secondary episodes of star formation, we can estimate the value of
$\Psi_{1/3}$ consistent with the observed scatter.  Applying this to the
ellipticals in Coma and Virgo for which the observed scatter is
$\delta(U-V)$=0.04 mag,  Bower et al concluded that, at most, 10\% of the
stellar mass could have been created in secondary bursts in the previous 5
Gyr, i.e. $\Psi_{1/2}<$0.1. In a more recent analysis however Bower et al
(1998) concluded that such constraints are rather weak when applied locally
particularly in truncated star formation histories. 

Noting these uncertainties will be reduced at larger look back times where we
are closer to the initial burst, a direct application of the same methodology
for the Ellis et al z=0.55 cluster sample yields $\Psi_{1/3}<$0.08 rms for the
cluster population. As a working definition, therefore, we will therefore
assume that $\Psi_{1/3}<$0.05 corresponds to a single-burst elliptical which
formed all its stars at a formation redshift $z=z_F$. Given the mean redshift
of our E/S0 sample is $\overline{z}\simeq$0.7, in an Einstein-de Sitter
universe with $H_o$=70 kms s$^{-1}$ Mpc$^{-1}$, this constraint essentially
implies a constant comoving density of ellipticals to $z\simeq$=1.5 for
$z_F\simeq$3. For $\Omega$=0.05, the constraint is only slightly weaker
($z\simeq$1.3). In the monolithic collapse model, we would not expect any of
our field ellipticals to have $\Psi_{1/3}>$0.05.

Examining Table 2, we can see that 4/11 of the faint E/S0s in the sample show
signs of recent star-formation according to the above definitions.  This
implies the majority, though not all, of the field E/S0s in the HDF completed
the bulk (95\%) of their star formation before $z\simeq$1-1.5. There is no
obvious trend with the galaxies which show evidence for recent star formation.
HDF-2.251 is in this category and is a strong radio source, but so is
HDF-4.752 which has a low $\Psi_{1/3}$. Where we believe the ellipticals can
be distinguished from the S0s, there is no correlation with $\Psi_{1/3}$.

It seems most unlikely that those few spheroidals which show a colour
dispersion could be explained as old systems with metallicity gradients.  As
Figure 1 shows, even extreme metallicity variations would be insufficient to
account for the colour variations, and in any case such variations would have
to operate in the direction of {\em depleting} the core of the elliptical,
which is opposite to local observations which find the cores of ellipticals to
be relatively enriched.

The robustness of this result, which suffers of course from the small sample
size in the tiny field of the HDF, can be considered by examining $\langle
{\cal L} \rangle$, the mean likelihood for all the pixels in the fit. Tests
indicate a value above 0.6 is a reasonably good fit to the colour
distribution.  Table 2 also lists $\langle E_{B-V}\rangle$, the mean $E_{B-V}$
reddening in the rest frame across all the pixels. Even allowing dust to be a
free parameter, the only elliptical which shows any evidence for appreciable
dust signature is the most distant example, HDF-4.928.

Does the large fraction (7/11) of galaxies with $\Psi_{1/3}<$0.05 contradict
the claim (Kauffmann et al 1996) that by a redshift of 1, the comoving volume
density of passively evolving red ellipticals should have declined by a factor
of 3? A detailed comparison must await more data but it is interesting to
consider the matter briefly notwithstanding the uncertainties. Semi-analytical
models do predict a substantial amount of merging at late times. A preliminary
inspection of the merging histories of massive field ellipticals at
$z\simeq$0.7 suggests somewhat higher percentages of star formation in the
previous few Gyr than indicated in Table 2 (Cole, private communication).

At face value therefore, there appears to be limited evidence for a more
diverse star formation history for HDF field ellipticals compared to their
rich cluster counterparts at similar redshift and, to the extent that
comparisons with hierarchical models can be currently made, the scatter seems
somewhat less than predicted. However, an important point to recognise in
these comparisons is the different techniques used in the field and cluster
data and, more fundamentally, what is meant by an `elliptical' in the
numerical simulations. For example, a centrally-concentrated but asymmetric
galaxy with tidal tails would probably not be classified by an observer as an
early-type system, while such a system might be classed as an elliptical in a
semi-analytical model on the basis of bulge-to-disk ratio.

It is also worth emphasising that our analysis is really addressing the
history of ellipticals only from one particular viewpoint. Rather than
measuring the absolute number density of systems that remain passively red at
various redshifts (c.f. Kauffmann et al 1996), we are here examining the
history of recognisable {\it spheroidal products}. It seems, from the number
counts in the HDF that these represent a substantial fraction of the present
day population of E/S0s. What we find is that only a few of these show
evidence of recent mergers through signatures of star formation. Of course
this may be simply because to produce a smooth product recognisable by us as a
E/S0, the merger {\it had}, by definition, to have occurred at an epoch such
that $\Psi_{1/3}<$0.05.  The ultimate resolution of this problem would then
lie in the joint luminosity-colour distribution of a large sample of
morphologically-selected field spheroidals (Menanteau et al, in preparation).

%%% End of new bit added by RSE

\begin{figure*}
\centering
\epsfig{figure=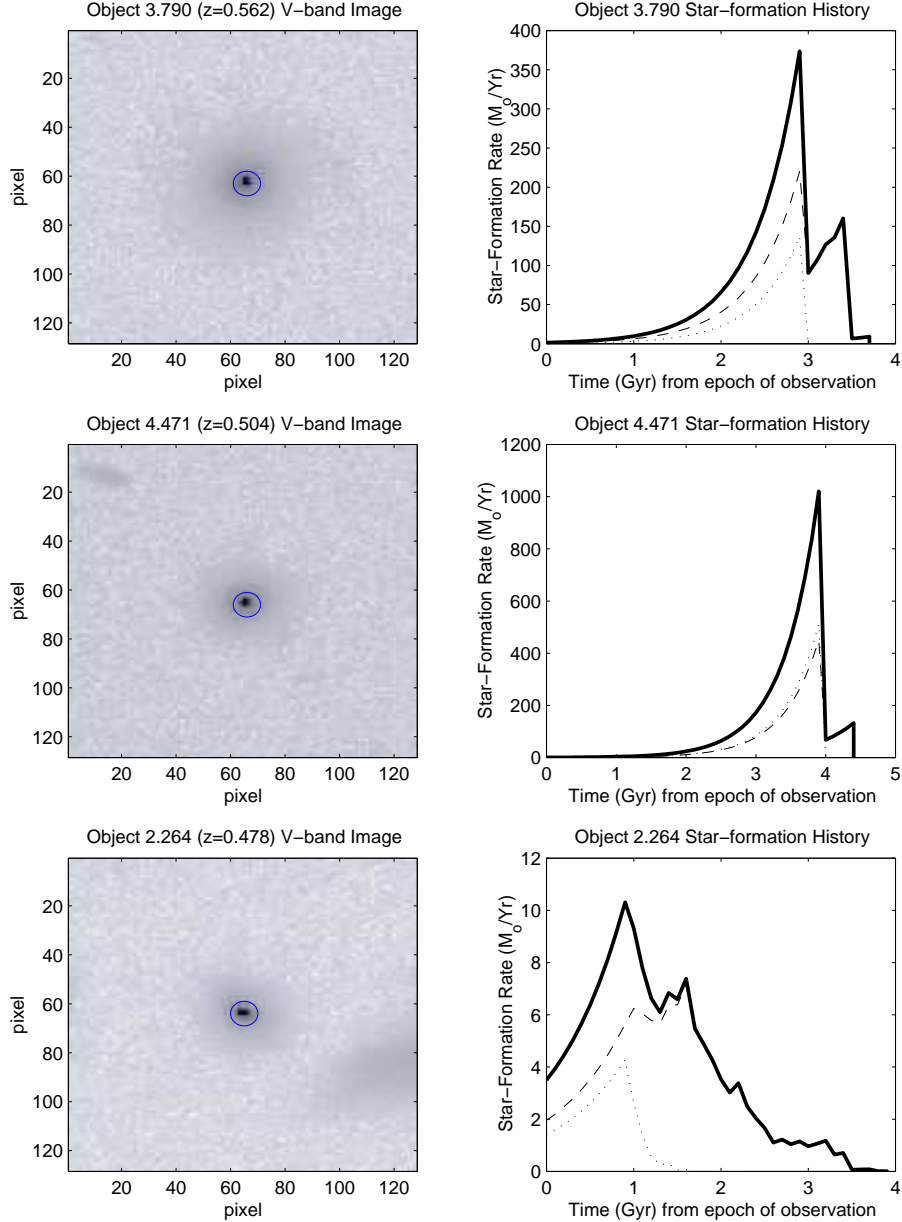,width=120truemm}
\caption{\it V-band images (left) and star formation histories (right)
for 3 spheroidal galaxies from the sample of 12 E/S0s from Bouwens et
al's list. The dotted curve on the right refers to the history
for pixels within the five pixel inner aperture shown on the left,
corresponding to the nuclear light. The dashed curve refers to the
light exterior to this aperture, and the solid curve refers to
the entire galaxy.
Models assume solar metallicity and no reddening.}
\end{figure*}

\subsection{Disc and Bulge Histories}  

We now turn to discuss the sample of 13 spirals from Table 1. As noted
earlier, our classifications are in reasonably good agreement with those of
Bouwens et al. {\em except} for a number of systems that Bouwens et al.
classify as spirals which we class as morphologically peculiar. In order to
ensure that at least no intermediate-late spirals are missed in our analysis,
and recognizing that classification of morphological peculiarities is rather
subjective, Table 3 includes {\em all} objects classified by Bouwens et al. as
spirals, including several galaxies which we class as peculiar or merging.  As
usual, we excluded systems whose redshifts were below $z=0.3$.  For reference,
our own morphological classifications are shown as the final column in Table~3.

In the case of the spiral galaxies we are interested in addressing a number of
questions. The first issue (hitherto unaddressed in HST studies) relates to
the relative ages of bulge and disc components.  Figure 7 indicates the
relative star formation histories for light inside and outside a five pixel
radius aperture centred on each of three spiral galaxies in our sample.  In
this paper no attempt has been made to do a photometric decomposition on these
images, but throughout the remainder of this paper we will refer to the the
light interior to the aperture as the `bulge', and exterior to the aperture as
the `disc', both for convenience and because the 5 pixel radius was chosen so
that bulge light clearly dominates the interior of the aperture for all
galaxies with an obvious bulge in the I-band images.  It is clear that the
inner and outer parts of the galaxies in Figure~7 have undergone remarkably
different star-formation histories. The ratio of the {\em median} ages of the
bulge and disc light, $T_{B/D}$, is given in Table 3 for various dust and
metallicity assumptions. We have included irregular and peculiar systems in
this tabulation, as well as spirals. As with the earlier statistic used to
quantify the star-formation history of ellipticals, the $T_{B/D}$ statistic,
which is a measure of {\em relative} age, is remarkably robust to changes in
metallicity and dust content. In the case of the 3 illustrated spirals, a mean
ratio of 1.54 emerges indicating the bulges (more accurately, the inner parts
of the galaxies) are 50\% older than the outer parts with little recent
activity. This ratio is somewhat larger than the mean for the entire sample,
but the general result that clearly emerges is that median age of bulges is
substantially older than the median ages of discs. In Table 3 we delineate
systems as having an `old bulge' if $T_{B/D}>1.1$ for {\em all three} model
scenarios, ie. requiring the inner light to be 10\% older than the outer light
regardless of assumptions made with regard to metallicity or dust content. Six
of the nine morphologically normal spirals pass this very stringent test. A
further two have $T_{B/D}>1$ for all three models, but with the relative age
difference below 10\%. Only a single spiral has $T_{B/D}<1$ for {\em any}
choice of metallicity or dust fraction, and even in that case the bulge is
only younger than the disk for one of the three model scenarios, and at a
negligible level (0.5\%).  The result that clearly emerges from our
investigation is that, for morphologically normal systems, {\em bulges are
indeed always the oldest parts of galaxies.} This remains true even in the
face of possible metallicity gradients, which are not modelled in this paper.
Based on observations in our own Galaxy and in the Local Group, the general
expectation is of course that bulges would have generally lower metallicity
than disks. This results in relatively {\em older} bulges, since generally
speaking decreasing metallicity results in increased age at a given colour, as
shown in Figure~1. As emphasized earlier, Table~3 shows this is an extremely
robust result that is independent of assumptions made with regard to
metallicity and dust.

We note that the bulges in the morphologically normal spirals considered here,
even in the oldest cases, do not appear to be as uniformly red and old as the
oldest ellipticals discussed in $\S$4.1.  This may be partly the result of
photometric contamination by underlying disk light (although the rather small
5 pixel radius aperture was chosen to minimize these effects), but it must
also be remembered that the mean look-back to the latter sample is 1-2 Gyr
higher.

For the four morphologically peculiar systems in Table~1, a rather different
picture emerges. Only one of these passes our stringent `old bulge' test, with
two of the remainder showing quite clear evidence for a systematically younger
bulge. The star-formation histories of morphologically peculiar galaxies,
including these systems (which are clear examples of the ``Blue Nucleated
Galaxies" reported in Schade et al. 1995), will be considered in detail in a
subsequent paper

The second substantial issue we wish to address with our analysis of
morphologically normal spirals, following the studies of Brinchmann et al
(1997) and Lilly et al (1997), is the suggestion from these papers that large
disc galaxies are well-established systems with a declining star formation
history. Brinchmann et al found the redshift distribution, $N(z)$, of the
CFRS/LDSS spirals was more extended than would be expected in a no-evolution
case. Very approximately, they concluded 1 mag of luminosity evolution would
be needed by $z\simeq$1 to reconcile predictions with the data. Lilly et al
(1997) analysed the surface photometry of the larger spirals and found a
similar effect.  Specifically, over z$<$0.7 they found a surface brightness
evolution of $\Delta\mu_B$=-0.8 mag. The trends are somewhat stronger than
those found in the dynamical study of Vogt et al (1997) although differences
in the way in which galaxies are selected can crucially affect the
interpretation, as Lilly et al discuss.

For the examples shown in Figure~7, our analysis indicates a
strongly-declining total star-formation rate with time and suggests a peak
activity 1-3 Gyr earlier. Over the entire sample this is a fairly consistent
trend. The mean redshift of the sample is $\overline{z}$=0.55, somewhat lower
than for the E/S0s, and thus the peak activity for the population would, in
most cases, occur at $z\simeq$1.

The bulk of the luminosity evolution inferred in the spiral population clearly
arises from the fading discs. We can therefore ask whether the trends seen for
the large face-on spirals in Figure 7 are consistent with the evolution in
blue light inferred from the CFRS/LDSS sample.  To examine this we took the
luminosity-weighted timescale $\tau$ (c.f. $\S$2.1) and mean age for
representative galaxies and estimated their rest-frame evolution in $M_B$. As
expected, the results depend much more sensitively on the input assumptions
but, in general, the results imply $\simeq$0.5 mag over 0.5$<z<$1 which is
consistent with the CFRS/LDSS constraints.

\begin{table*}
\centering
\begin{minipage}{170mm}
\caption{Spiral Galaxies: Ratio of Inner to Outer Median Age}
\begin{tabular}{@{}cccccccccccc}
\hline
      &     &   \multicolumn{3}{c}{solar + dust} &   \multicolumn{3}{c}{30\% solar + dust} & \multicolumn{3}{c}{solar + no dust} & \\
 ID   &  $z$  &  $T_{B/D}$ &  $\langle {\cal L} \rangle$\footnote{Mean likelihood for all pixels.}  & $\langle E_{B-V}\rangle$\footnote{Mean rest-frame relative $B - V$ dust reddening for all pixels.} &  $T_{B/D}$ &  $\langle {\cal L} \rangle$  & $\langle E_{B-V}\rangle$ &  $T_{B/D}$ &  $\langle {\cal L} \rangle$  & $\langle E_{B-V}\rangle$ &  comment \\
\hline\hline
3.610    &      0.517  &     1.490  &     0.829 &     0.084 &     1.212 &     0.884 &     0.105  &     1.783    &        0.804   &       0.000  &  Spiral -- old bulge \\
4.656    &      0.454  &     1.400  &     0.919 &     0.063 &     1.214 &     0.791 &     0.027  &     1.154    &        0.838   &       0.000  &  Spiral -- old bulge \\
3.350    &      0.642  &     1.231  &     0.947 &     0.094 &     1.316 &     0.925 &     0.047  &     1.200    &        0.866   &       0.000  &  Spiral -- old bulge \\
4.795    &      0.432  &     1.294  &     0.844 &     0.040 &     1.727 &     0.691 &     0.016  &     1.421    &        0.816   &       0.000  &  Spiral -- old bulge \\
4.550    &      1.012  &     1.222  &     0.915 &     0.065 &     1.185 &     0.863 &     0.077  &     1.290    &        0.882   &       0.000  &  Spiral -- old bulge \\
3.534    &      0.319  &     1.571  &     0.926 &     0.031 &     1.667 &     0.860 &     0.025  &     1.471    &        0.838   &       0.000  &  Barred Spiral -- old bulge \\
3.400    &      0.473  &     1.000  &     0.904 &     0.056 &     0.625 &     0.867 &     0.051  &     0.875    &        0.865   &       0.000  &  Peculiar           \\
3.264    &      0.475  &     1.308  &     0.902 &     0.103 &     1.261 &     0.924 &     0.080  &     1.095    &        0.769   &       0.000  &  Spiral           \\
3.659    &      0.299  &     1.115  &     0.705 &     0.048 &     1.167 &     0.696 &     0.044  &     1.067    &        0.685   &       0.000  &  Sab           \\
4.402    &      0.555  &     0.400  &     0.762 &     0.039 &     0.500 &     0.682 &     0.030  &     0.500    &        0.728   &       0.000  &  Spec           \\
3.486    &      0.790  &     1.643  &     0.899 &     0.085 &     1.833 &     0.857 &     0.071  &     1.400    &        0.865   &       0.000  &  Spiral -- old bulge \\
3.143    &      0.475  &     1.467  &     0.945 &     0.050 &     1.250 &     0.831 &     0.012  &     1.250    &        0.900   &       0.000  &  Peculiar -- old bulge \\
4.950    &      0.608  &     1.000  &     0.873 &     0.013 &     1.000 &     0.739 &     0.002  &     1.167    &        0.857   &       0.000  &  Irregular           \\
\hline
\end{tabular}
\end{minipage}
\end{table*}

\begin{figure*}
\centering
\epsfig{figure=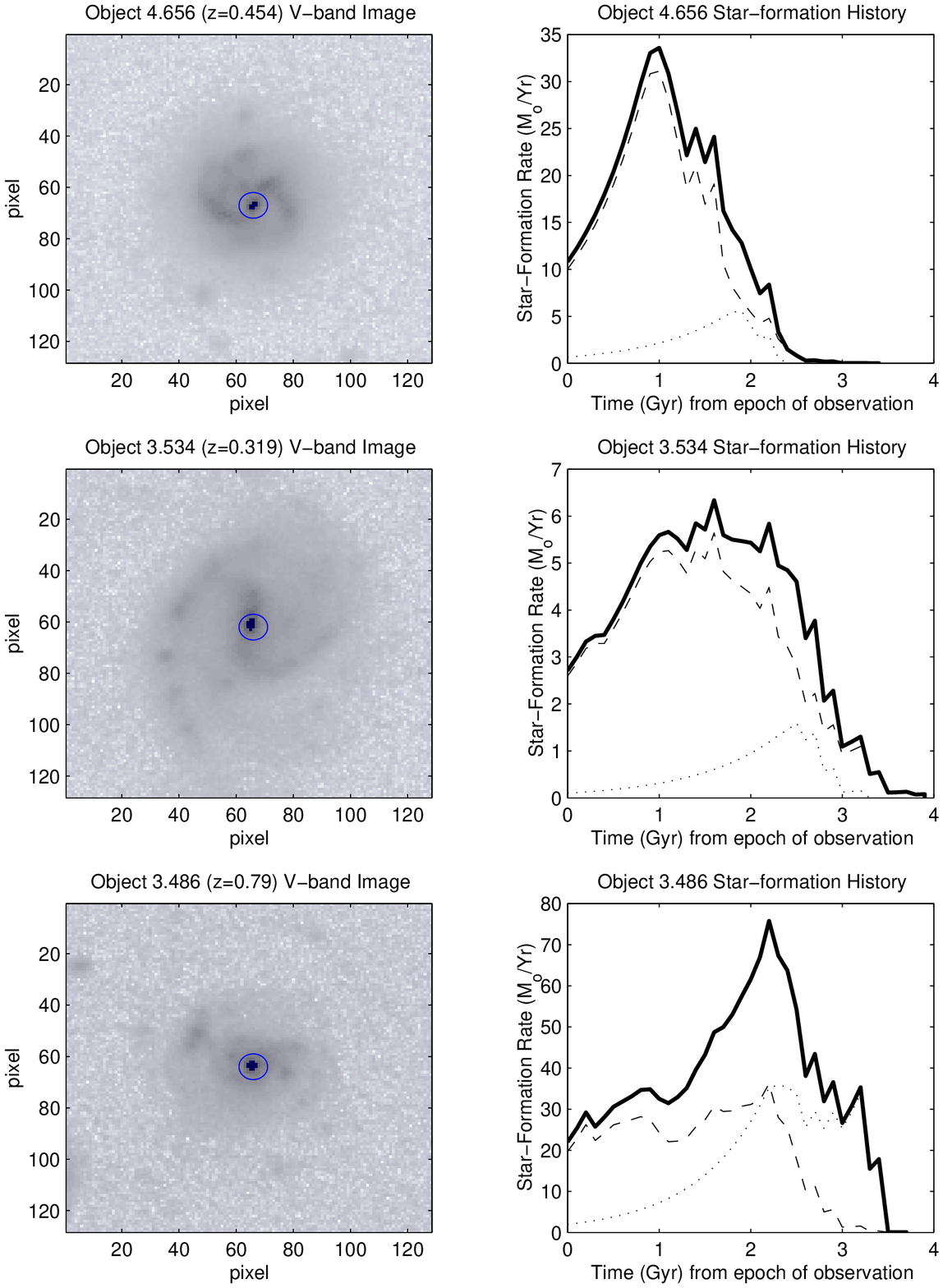,width=120truemm}
\caption{\it As for Figure 6, for a sample of spirals from the Bouwens et
al sample. The dotted curve corresponds to the aperture shown on the
left, which isolates bulge light. The dashed curve corresponds to
pixels outside this aperture, and the solid curve is the history
for the entire galaxy.}
\end{figure*}

\subsection{Global Histories}

The detailed analysis of the colour distribution of individual galaxies and
those of a given morphology has given a good indication of the advantages and
restrictions of our method. However, if we wish to ultimately derive
conclusions concerning the star formation history of volume-averaged
populations, we need to extend the techniques taking into account the
detectability of each galaxy as a function of $z$ and the relative masses
involved in the various star formation phases.  This is an important point
notwithstanding the conclusions in $\S$4.1-4.2 since, as various authors have
discussed (White 1996, Brinchmann et al 1997) morphology may be a transient
property. In principle our method offers an opportunity to examine more
general trends -- as discussed in the next section. A major conceptual
advantage of our method is that by focussing on star-formation history on a
pixel-by-pixel basis, we probe galaxian star-formation history in a manner
that is meaningful even in the face of morphological transformations. Even in
an evolutionary scenario dominated by density evolution, stars from the merger
fragments are still present in the merger product. So the structure of the
star-formation history curves shown in Figure 7 is a partial record of the
formation history of the stars in the final system, even if we have no
information about the morphologies of the fragments.  We can thus consider
whether the Bouwens et al galaxy sample is predominantly the result of
star-formation activity which peaked at $z\simeq$1-2, as claimed by Madau et
al 1998.

There are several complications in conducting such calculations.  Firstly,
taking this approach requires us to assume {\em absolute} ages for the galaxy
sample, rather than the more robust relative ages for internal galaxian
components, upon which we have focussed so far. An equally important
complication is that the Bouwens et al sample is flux-limited and so spans a
range in look-back time and luminosities.  To convert discrete realisations of
the SF history $\Psi_i(t)$ for $i$=1,N objects spanning a range of redshifts
$z_i$, into its comoving average, $\Psi_T$, we must adopt a cosmological model
$t(z,H_o,\Omega)$, and weight each $i$th contribution by $1/V_{max}$ (Schmidt
1968; Bouwens et al. 1998):

\begin{equation}
$$\Psi_T(z) = \Sigma_{i=1}^N\, 1/V_{max} \Psi_i[z_i, z(t,H_o,\Omega)]$$
\end{equation}

As an illustration of this, we have attempted to combine the histories
$\Psi_i(t)$ for most of the galaxies in Table 1 in order to produce an
estimate of the global history $\Psi_T(z)$ (Figure 8) where we compare our
results with that of Madau et al (1997). We show curves as an envelope whose
limits correspond to 30\% solar and 150\% solar metallicity for the entire
sample. We cut off the diagram at $z=1.5$ (with the last observational data
point corresponding to $z = 1.2$ because of the small sample size considered
-- at $z\sim1.5$ one or two individual bright galaxies dominate the total
star-formation history plot, and Poisson errors become dominant.  The curves
correspond to the volume-averaged star-formation histories for all spirals,
peculiars, irregulars, and the four ``young'' ellipticals in the sample.
``Coeval'' ellipticals have been  excluded from consideration because the
contribution at high redshift from these systems depends crucially on their
adopted absolute ages (which are very poorly constrained because, by
definition, they have little dispersion in internal colour). In any case these
systems are likely to contribute mostly at $z>1.2$.

Although clearly an approximation (whose correctness is subject to the
necessity for assuming absolute, rather than relative, ages in the
calculation), the extrapolated star-formation history of the Universe
calculated by adding together results from just $\sim 20$ galaxies in the
Bouwens sample approximates that obtained completely independently from deep
galaxy surveys. At redshifts $z>0.5$ the survey data points lie slightly below
the predicted curve from our analysis, but this may not be significant given
the rather large uncertainties on both the models and the data points, and
given the omission of ``old ellipticals'' (which must be contributing a small
amount of star formation even at $z<1.2$) from the models.

At face value Figure~8 suggests that roughly all the starlight located in deep
surveys at $z<1.2$ ultimately ends up in galaxies represented by the systems
in the Bouwens sample. While interesting, this conclusion is highly uncertain,
not least because of the very small sample size used in this extrapolation. In
particular, we have already noted there is an apparent deficiency of
irregular/peculiar/merger galaxies in the Bouwens et al sample compared to
wider field HST studies. If the $z<$1 inventory is richer in star forming
systems whose evolution is important (as suggested by the results of
Brinchmann et al) then either the contribution of late type systems beyond
$z\simeq$1 cannot be dominant or the extrapolation is unreliable. An obvious
example would be the presence of dust, not in the currently-observed Bouwens
et al sample or additional sources thought to be under-represented at $z<$1,
but perhaps in their predecessors seen at earlier times. At this stage, this
is largely speculation. The major result is the predictive power of our
colour-based techniques particularly when enhanced with larger and deeper
samples.

%In other words, the final inventory of
%stars seems to add up in a consistent manner at all redshifts
%$0.3<z<1.2$ given the numbers of galaxies in the Bouwens sample and our
%reconstructed star-formation histories. This is an important
%conclusion, arguing (for example) that high-redshift low surface
%brightness systems or dust-enshrouded galaxies are {\em not} playing a
%dominant role in terms of the ultimate stellar content of the Universe.
%If either of these were true, then one of two (rather unlikely)
%alternatives would also have to hold true:  (a)~If luminosity evolution
%prevails then undetected systems at high redshift must remain invisible
%in roughly the same proportion at all redshifts, or (b) if density
%evolution dominates then the stars in undetected high-redshift galaxies
%cannot have ended up in lower redshift systems via mergers.

\begin{figure*} \centering \epsfig{figure=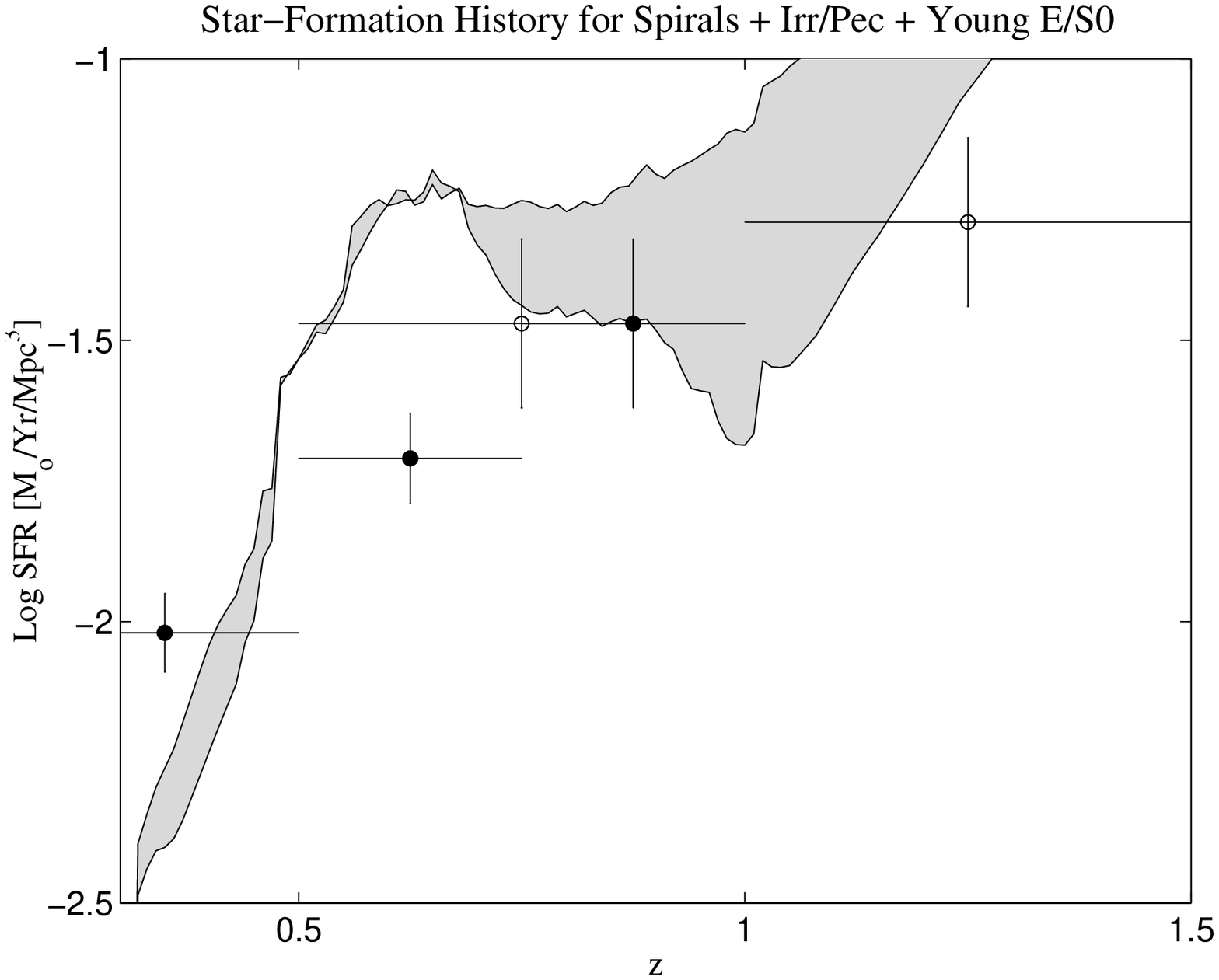,width=110mm} \caption{\it
The global star-formation history of the universe obtained by adding the
contribution from the spirals, irregulars and evolving ellipticals in our
sample. The recovered total star-formation history assuming 30\% solar
metallicity for all galaxies is shown as the top curve, with  the
corresponding curve for 150\% solar metallicity shown as the bottom curve. The
interior region (shaded gray) thus corresponds to the region of the diagram
possible for a mixture of metallicities.  Calzetti-law dust extinction was
assumed for all curves. The solid points correspond to data from the CFRS
survey (Lilly et al 1996), as given in Madau et al. (1996). Open circles
correspond to estimates based on photometric redshifts in the Hubble Deep
Field (Connolly et al 1997).} 
\end{figure*}

\section{Discussion}

It is interesting to contrast our method with the approach of Brinchmann et al
(1998) who derived luminosity functions and volume-averaged luminosity
densities for morphological subsets from the combination of HST imaging and
ground-based redshift surveys for $\simeq$300 field galaxies. Whilst they
discussed the validity of the assumption that the morphological type is a time
invariant property, their analysis was primarily concerned with calculating
the contribution of each class to the redshift-dependent star formation rate.
They found the greatest rate change with redshift over 0$<z<$1 belonged to
galaxies of irregular morphologies although the largest contribution to the
volume-averaged total at $z\simeq$1 was from disc galaxies. Brinchmann et al
argued that it is unlikely that the rapidly declining population of galaxies
with irregular morphologies transform to form large disc galaxies given the
contrast in evolutionary trends. 

Even with $>$300 galaxies with redshifts and HST morphologies, once
the sample is broken into redshift and morphology bins, the statistics
necessary to reach convincing conclusions becomes marginal.
Furthermore, the likelihood of a possible migration from one class to
another as the star formation rate changes or, more simply, a fading
below the detection limit as the rate declines could easily complicate
the above conclusions.

Our technique offers a number of ways with which to overcome the limitations
of the Brinchmann et al analysis. Foremost, we can address in more detail the
question of the homogeneity in the star-formation history of a given class.
The issue here is not simply the scatter from one galaxy to another within a
given class, but the dispersion in internal pixel colours and the implications
of this for the star-formation histories.  Admittedly there is no short cut to
having a fair and representative sample of galaxies to analyse, but the
availability of resolved data opens up a much more powerful tool for
addressing issues related, for example, to the question of whether most field
ellipticals are simple coeval stellar populations. The second advantage of our
method is that we can, within the limitations discussed earlier, by focussing
on star-formation sidestep the difficulty of not knowing the precursor
population of each class.  Indeed, in principle, we can abandon morphological
categories altogether and simply isolate those $z<$1 galaxies which are likely
to have dominated the star formation history 2-5 Gyr earlier. Finally, of
course, we can address the history of distinct sub-components such as bulges
and discs.

It is important to consider ways to verify the validity of many of our
conclusions. Concerning the issue of whether spheroidal galaxies are being
continuously produced, our approach in examining evidence in the colours for
recent mergers can be easily extended to higher redshifts.  The technique
complements other approaches based on colour distributions.  For the disc
galaxy evolution, it will be possible to verify our star formation rates
directly, and constrain dust content, once integral field spectroscopy becomes
feasible to the necessary surface brightness limits.  The greatest restriction
in the application of our method is the need for redshifts which, presently,
restricts the HDF sample to only 32 galaxies. By extending the HDF surveys to
deeper spectroscopic limits or perhaps adopting photometric redshifts, this
restriction can be partially overcome. A greater redshift range would allow
intermediate redshift samples to be directly compared with their statistical
ancestors for consistency purposes.

A second enhancement is the extension of our method to near-infrared
passbands. Without infrared observations to help constrain the appropriate
dust law, it is difficult to constrain definitively the possible interplay
between dust and genuine stellar histories. Even with ground-based photometry,
the integrated near-infrared magnitudes offer a way to verify the global
properties inferred for a given galaxy, but ultimately we can expect to use
NICMOS data to extend the full resolved colour-colour analysis.  Later papers
in the series will take these ideas forward and address more closely the role
of the star forming irregulars thought to be dominant in the intermediate
redshift range. 

\section{Conclusions}

We summarise our conclusions as follows:
\vspace{1cm}

1. The spectral synthesis modelling of the spatially resolved internal colours
of galaxies of known redshift is an important new technique for probing galaxy
evolution, offering a number of benefits compared to studies based on
integrated colour alone. The technique makes an explicit connection between
evolutionary history and stellar populations. We use the dispersion around
colour-colour tracks to distinguish between models that are degenerate in
terms of integrated colour enabling the dust content to be explicitly
parameterized and allowing the star-formation histories of physically-distinct
galactic subcomponents to be reconstructed. Most importantly, internal colour
analyses enable robust studies of {\em relative ages} addressing the
homogeneity of stellar populations and probing the order in which galactic
components have been assembled.

2. Most early-type systems (7/11) in the Bouwens et al. redshift sample have
internal colour dispersions consistent with being old and coeval, but
(adopting $\Psi_{1/3}<0.05$ as the criterion for an old, coeval system) {\em
over one-third [4/11] of the early type systems studied do not}.  This result
contrasts with that derived from the scatter in colour-magnitude diagrams for
populations of spheroidals in distant rich clusters. The field sample is much
too small for detailed conclusions to be drawn regarding the star-formation
history of early type systems as a class, but the current fraction ($\sim
40\%$) of young early-type systems is reasonably consistent with the
predictions from hierarchical models. If one adopts a less stringent
definition for old, coeval system (say $\Psi_{1/3}<0.1$) then the constraints
on hierarchical models become substantially tighter. We therefore suggest that
specific predictions for the distribution of $\Psi_{1/3}$ would provide a
valuable test of the success of hierarchical models.  However, quantifying the
fraction of young early-type galaxies expected in such scenarios depends on
how quickly such systems relax after a merger event, and hence how such
systems are classified both visually and on the basis of n-body/semi-analytic
models. Similar criteria must be adopted for both observations and models.

3. We find that the bulges in the HDF spirals are generally redder than their
disks with a median age unambiguously older in the majority (6/9) and at least
as old as the disk in most of the remainder (2/9). Bulges exhibit
qualitatively different star-formation histories from their surrounding disks
indicating a short initial period of star-formation followed by relative
quiescence. Unless bulges are heavily enriched relative to disks, these
results strongly support the ``traditional'' picture for the formation of
bulges and pose a serious challenge to competing secular evolution models.

4. The overall dust content in galaxies from the Bouwens et al sample at
intermediate redshifts $0.3<z<1.0$ is fairly similar to that seen locally.
Spirals and irregulars exhibit evidence for patchy pockets of dust at the
$E_{B-V}\sim0.1$ level, but the majority (10/11) of early-type galaxies appear
essentially dust-free. No galaxy in this sample appears to have internal
colour dispersions suggestive of major obscuration nor internal colours
dominated by dust as opposed to age or star-formation history. As in the local
Universe, dust in the intermediate-redshift Universe ($z<1$) appears to play a
subsidiary role in the interpretation of galaxy morphology and number counts.
Our conclusions are dependent, however, on the universal validity of the
Calzetti (1997) extinction law and independent tests of the applicability of
this law over a range of redshifts and galaxy types are of the first
importance.

5. The star-formation histories we infer for field spirals agree with those
derived from deep galaxy surveys. If significant morphological transformations
have not occurred within our spiral sample, then our declining star-formation
histories are consistent with typical fading in disk light by $M_B \sim 0.4$
mag between $z=1$ and $z=0.5$, in agreement with the values obtained from the
CFRS and LDSS surveys (Brinchmann et al. 1998; Lilly et al.  1998). The peak
in the volume-averaged star-formation activity contributed by starlight in $z
\sim 0.5$ disc galaxies occurred close to $z=1$.

6. As our methodology focusses on the stellar content (rather than on the bulk
morphological properties), we illustrate how an approximate inventory of the
total starlight in our galaxy sample can be backtracked towards the formation
epoch without assuming that morphologies are preserved. Surprisingly, we
recover the bulk of the starlight seen in deeper surveys. The present sample
is very small and this is a tentative result, particularly as there is some
evidence that the Bouwens et al sample is deficient in star forming galaxies.
However, if confirmed, it would suggest that star-formation at $z<1.2$ is not
occurring dominantly in systems which have escaped detection in deep optical
surveys.

{\bf Acknowledgments} 

We thank Simon Lilly, Simon White, Sidney van den Bergh, Max Pettini,
Jarle Brinchmann, Rychard Bouwens, and Felipe Menanteau for useful 
discussions and suggestions.

\end{document}